# Superconducting non-volatile memory based on charge trapping and gate-controlled supercurrent

L. Ruf[1*], A. Di Bernardo[1*], E. Scheer[1*]

1 . *Department of Physics, University of Konstanz, Universitätsstraße 10, 78464 Konstanz, Germany.*

Email(s): leon.ruf@uni-konstanz.de
angelo.dibernardo@uni-konstanz.de
elke.scheer@uni-konstanz.de

## Abstract

Superconducting electronics holds great promise for energy-efficient high-performance and quantum computing, yet no superconducting memory has matched the performance of conventional semiconductor memories – a long-standing bottleneck. Here we demonstrate a voltage-controlled, non-volatile superconducting memory that exploits two previously independent effects: gate-controlled supercurrent (GCS), the gate-voltage-induced suppression of the critical current $I_c$ in a superconducting constriction, and charge trapping in an $Al_2O_3$ dielectric. Trapped charges shift the threshold gate voltage required for $I_c$ suppression, defining two stable, well-separated $I_c$ states that can be used to store binary information. We demonstrate reliable non-destructive readout and reversible write/erase cycling over nearly fifty consecutive cycles with the device remaining in the zero-resistance state throughout. Stored information survives thermal cycling well above the superconducting transition temperature $T_c$, confirming true non-volatility – a capability absent in all existing superconducting memories. We further discuss integration into a NAND architecture and show significant power-dissipation advantages over CMOS charge-trap flash memories.

## Teaser

Gate-controlled supercurrent combined with charge trapping realizes a non-volatile superconducting memory.

## INTRODUCTION

The rapid advancement of artificial intelligence is driving the demand for high-performance computing (HPC) based on complementary metal-oxide-semiconductor (CMOS) technology, which remains unmatched in scalability and cost-effectiveness (*1*). However, power dissipation and thermal management in CMOS systems pose fundamental challenges for the continued expansion of HPC (*2*).

Superconducting devices, which operate at cryogenic temperatures with near-zero energy dissipation, offer a compelling route to more energy-efficient HPC. While superconducting logic elements are already commercially available, the lack of a superconducting memory with performance comparable to CMOS memories remains a critical bottleneck toward the realization of fully superconducting HPC systems.

Existing superconducting memory concepts include Josephson junction-based devices (*3,4*) and superconducting quantum interference devices (SQUIDs) (*5*) which rely on persistent superconducting currents (supercurrents) or trapped magnetic flux to store information. Although fast and low-power (*6*), these devices suffer from sensitivity to external magnetic fields, limited scalability and the need for persistent supercurrents to retain information (*6,7*) – which restricts their operation to a temperature ($T$) below the superconducting transition temperature ($T_c$). Moreover, their current-controlled operation also complicates integration with voltage-controlled CMOS circuitry (*8*).

More recently, voltage-controlled devices like Josephson field-effect transistors (JoFETs) (*9,10*) and Josephson diodes (*11-14*) have also been explored. However, these devices require active biasing, and do not inherently support non-volatile memory operations. Alternative approaches including ferroelectric-gated JoFETs (*15*) and electrothermally-switched superconducting nanowires have recently demonstrated memory functionality (*16*) but involve either dissipative states or destructive memory readout.

In parallel, gate-controlled supercurrent (GCS) in metallic superconductors (Ss) has emerged as a versatile voltage-controlled phenomenon, whereby the application of a gate voltage ($V_G$) to an electrode near a nanoscale superconducting constriction suppresses its maximum switching current $I_c$ (*17-30*). While the microscopic origin of the GCS remains debated – with $V_G$-induced leakage current ($I_{leak}$) playing a central role in most cases – the effect is robust across different superconducting materials (*17,20,25,26*), substrates (*29*) and device geometries (*17,20,23,24,26,27*), which makes it attractive for superconducting electronics applications. To date, however, the GCS has been exploited almost exclusively for logic operations, while its potential for memory has remained largely unexplored.

Here, we demonstrate a new operational paradigm in which the GCS and charge trapping in an oxide dielectric (*31*) – two phenomena previously studied independently – are combined within a single device to realize a *non-volatile voltage-controlled superconducting memory.*

Through measurements on multiple devices, we show that $V_G$-induced charge trapping in a sapphire ($Al_2O_3$) dielectric substrate – a mechanism well-established in CMOS charge-trap memories (*32-35*) – reversibly shifts the threshold $V_G$ needed for the GCS, enabling the memory device to operate between two stable states associated with well-separated $I_c$ and $I_{leak}$ levels. These states can be repeatedly written or erased using $V_G$ values above a certain threshold $V_G$ and read non-destructively at a $V_G$ below the threshold while maintaining the device in the superconducting state. The stored information is robust against large $T$ excursions well above the $T_c$ of the device, demonstrating true non-volatility. Moreover, since information is stored in trapped charges rather than supercurrents, writing can be performed even in the normal (resistive) state above $T_c$. Finally, we discuss integration of these devices into a NAND architecture and show advantages over state-of-the-art CMOS NAND charge-trap layouts, particularly regarding reduced power dissipation during readout.

## RESULTS AND DISCUSSION

### Charge trapping in GCS devices with an $Al_2O_3$ dielectric

To investigate the effect of charge trapping on GCS devices, we fabricated and characterized nine devices (Devices A-I) consisting of three-terminal Nb Dayem bridges on an $Al_2O_3$ dielectric substrate. A false-color scanning electron micrograph of a representative device with the measurement configuration is shown in Fig. 1A.

For each device, at a fixed $T$ below $T_c$, we measured the $I_c$ from voltage-current, $V(I)$, characteristics (fig. S1, Devices A-D), and recorded both $I_c$ and $I_{leak}$ as a function of the effective gate voltage $V_G = V_S - I \cdot R_{wiring}$ (*18*), where $V_S$ is the sourced gate voltage and $I \cdot R_{wiring}$ accounts for voltage drops in the measurement circuit (Fig. 1, A and B). Device parameters including measurement $T$, $T_c$, $I_c$ in the absence of $V_G$ ($I_{c0}$), normal-state resistance ($R_N$) and other geometry factors are provided in Table S1.

Across all devices tested, $I_c$ decreases and $I_{leak}$ increases with increasing $|V_G|$, with these two quantities becoming anticorrelated after the onset of the $I_c$ suppression – consistent with the GCS effect and in agreement with previous reports (*18-27,30*). However, unlike previous studies, we find that both $I_c(V_G)$ and $I_{leak}(V_G)$ depend on the history of the applied $V_G$. We attribute this history dependence to $V_G$-induced charge trapping and detrapping in the $Al_2O_3$ substrate, which modifies the effective local capacitance $C_{eff}$ between the gate and the superconducting wire.

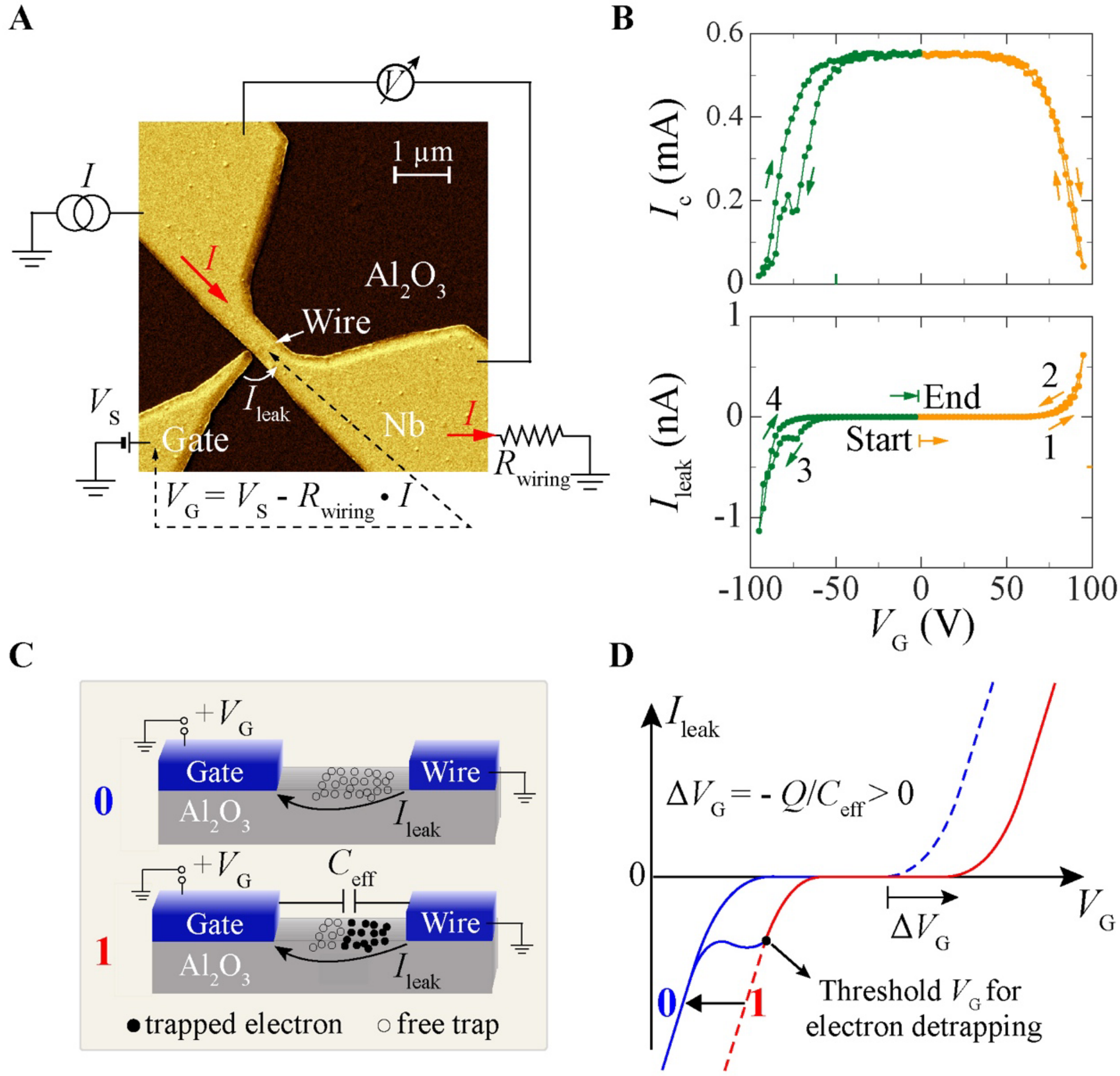

**Fig. 1. Charge trapping in Nb GCS devices on $Al_2O_3$.** (**A**) False-color scanning electron micrograph of a Nb Dayem bridge on an $Al_2O_3$ substrate with the four-point measurement configuration used to determine the critical current, $I_c$. The effective gate voltage $V_G = V_S - I{\cdot}R_{wiring}$ is applied between the gate and wire, where $V_S$ is the sourced gate voltage, $I$ the bias current and $R_{wiring}$ the total wiring resistance. The leakage current $I_{leak}$ is recorded simultaneously in a two-point configuration. (**B**) $I_c$ (top) and $I_{leak}$ (bottom) versus $V_G$ for Device G over a complete $V_G$ cycle starting from the virgin state ($V_S = 0$). $V_G$ was swept to a maximum positive value (curve 1, orange), back to zero (curve 2, orange), then to a maximum negative value (curve 3, green) and back to zero (curve 4, green); arrows indicate the sweep direction. Hysteresis between curves 3 and 4 and the non-monotonic kink in curve 3 are signatures of charge trapping. (**C** and **D**) Schematic cross section (side view) illustrating two charge-trapping states – state "0" (no trapped charge, blue) and state "1" (trapped electrons, red) – and their effect on $I_{leak}(V_G)$. Electron trapping (assumed in (D)) shifts $I_{leak}(V_G)$ toward higher $V_G$ by $\Delta V_G = -Q/C_{eff}$, where $Q$ is the net trapped charge (negative for electrons) and $C_{eff}$ is the effective gate-to-wire capacitance. Trapping and detrapping only occur above a certain threshold $V_G$ (D). Since $I_{leak}(V_G)$ and $I_c(V_G)$ are anticorrelated (B), the same shift also applies to $I_c(V_G)$.

Figure 1B shows $I_c(V_G)$ (top panel) and $I_{leak}(V_G)$ (bottom panel) measured over a complete $V_G$ cycle for Device G, starting from its virgin state (no prior $V_G$ applied). During the initial positive sweep (Fig. 1B, curve 1), increasing $V_G$ induces electron trapping in $Al_2O_3$ substrate near the superconducting wire, thus increasing $C_{eff}$ and shifting both $I_{leak}(V_G)$ and $I_c(V_G)$ toward higher $V_G$ (Fig. 1, C and D).

During the subsequent reverse sweep (Fig. 1B, curve 2), most electrons remain trapped, and both $I_c(V_G)$ and $I_{leak}(V_G)$ curves largely retrace the forward branch, confirming the stability of the trapped state. Sweeping then to negative $V_G$ (Fig. 1B, curve 3) reveals the stored charge as a shift of the onset of $I_c$ suppression and $I_{leak}$ increase toward smaller $|V_G|$ compared to the initial forward sweep. At sufficiently large $|V_G|$, electron detrapping – and possibly hole trapping – produces a non-monotonic kink, which marks a threshold $V_G$ for charge trapping and shifts both characteristics further toward negative $V_G$. Upon detrapping, $C_{eff}$ decreases, shifting the curves back and closing the hysteresis loop during the subsequent positive $V_G$ sweep (Fig. 1B, curve 4).

This hysteresis defines two stable and reproducible memory states (Fig. 1C): state "0" (blue, no trapped charge) and state "1" (red, electrons trapped), each associated with a distinct pair of $I_{leak}(V_G)$ and $I_c(V_G)$ curves (Fig. 1D). The $V_G$ shift, $\Delta V_G$, between states is set by the net trapped charge $Q$ and $C_{eff}$, with $\Delta V_G = -Q/C_{eff}$ (Fig. 1D). Electron trapping ($Q < 0$) results in a positive $\Delta V_G$, as shown in Figure 1.

**Effect of charge trapping on the operating voltage of GCS devices**

To further quantify the role of charge trapping in GCS devices, we performed measurements on Device C using alternating $V_G$ polarity (Fig. 2, A and B). Starting from the virgin state of the device, we applied repeated $V_G$ cycles, with each cycle consisting of a sweep from zero to a maximum positive value ($\tilde{V}_G$), followed by a sweep from 0 to -$\tilde{V}_G$. Unlike the continuous bidirectional cycle in Fig. 1B, both sweeps here start from $V_G = 0$, which allows the effects of positive and negative $V_G$ to be probed independently. Three such cycles were performed yielding six consecutive sweeps (labelled 1-6).

Figure 2A shows the time evolution of $V_G(t)$ (top panel), $|I_{leak}(t)|$ (middle panel), and $I_c(t)$ (bottom panel). Clear asymmetries in $|I_{leak}|(t)$ and $I_c(t)$ appear between positive (orange shading) and negative sweeps (green shading): positive sweeps following negative sweeps (sweep 3 and 5) exhibit non-monotonic response, whereas negative sweeps remain monotonic throughout. These non-monotonic trends are signatures of charge-trapping events, and manifest in $I_{leak}$ and $I_c$ – both in their time-dependent (Fig. 2A) and $V_G$-dependent (Fig. 2B) traces. We note that the slope changes visible in $V_G(t)$ (Fig. 2A, top panel) arise from the finer voltage step size used in the range where $I_c$ suppression occurs and are therefore unrelated to charge trapping.

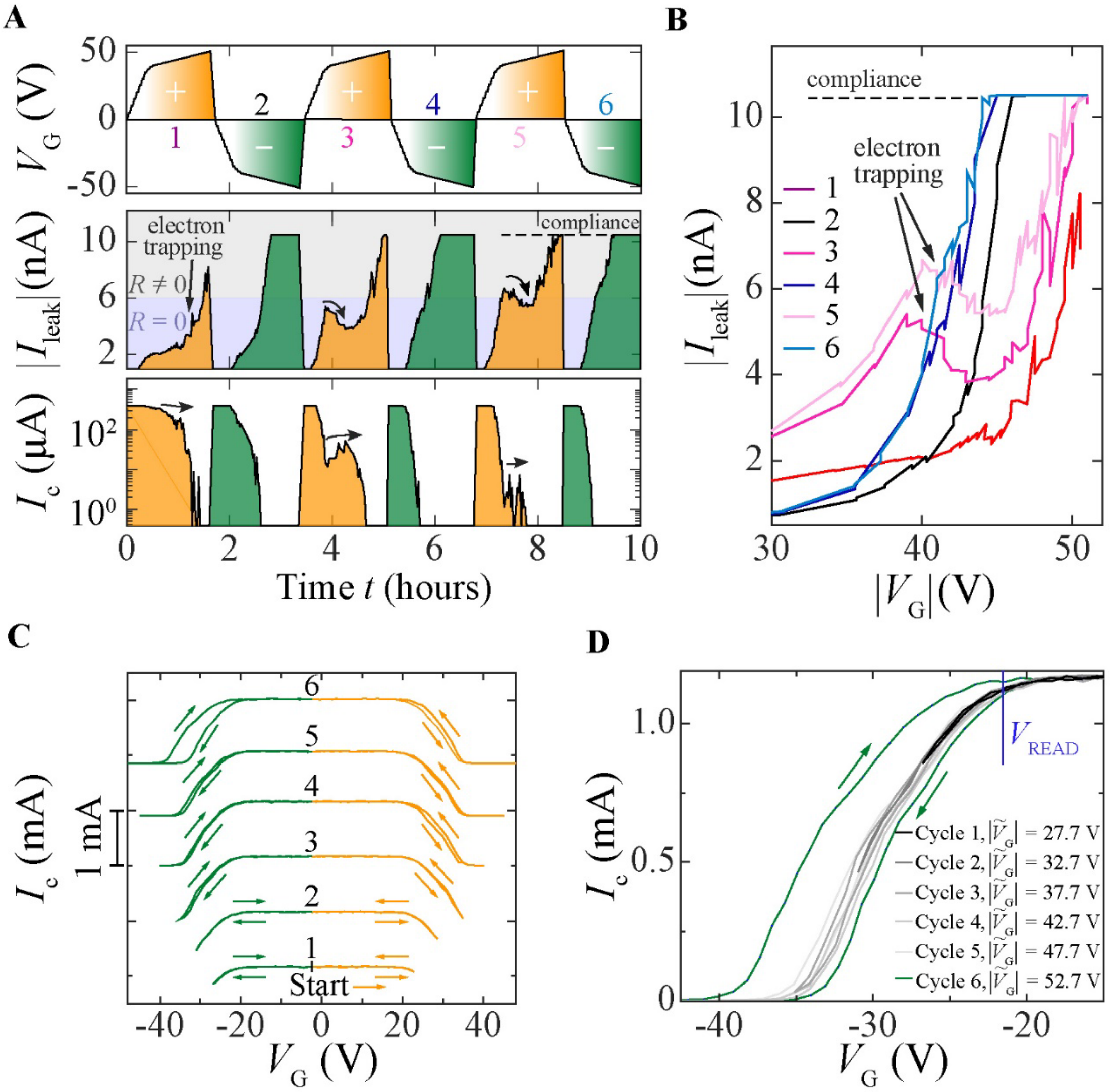

**Fig. 2. Shift in the operating voltage of a GCS device induced by charge trapping.** (**A**) Time traces of gate voltage $V_G(t)$ (top panel), leakage current $|I_{leak}|(t)$ (middle panel) and critical current $I_c(t)$ (bottom panel) for Device C, measured over six consecutive $V_G$ sweeps (labelled 1-6) alternating between positive (orange shading) and negative (green shading) $V_G$ polarity, each starting from $V_G = 0$. (**B**) $|I_{leak}|$ versus $|V_G|$ for the six $V_G$ sweeps in (A). Positive-polarity sweeps (1, 3, 5) are progressively shifted toward higher $|V_G|$ relative to negative-polarity sweeps and exhibit a non-monotonic kink (black arrows) – both signatures of charge trapping. (**C**) $I_c(V_G)$ for six consecutive $V_G$ cycles on Device F, initialized in the electron-trapped state (state "1", Fig. 1C), by a 500 ms $V_G$ pulse of +50.8 V (fig. S3). The amplitude of the maximum applied $V_G$, $|\tilde{V}_G|$, is incrementally increased after each cycle; traces are vertically offset by 1 mA for clarity. (**D**) $I_c(V_G)$ data from (C) restricted to negative $V_G$ and overlaid for direct comparison, with each curve corresponding to a different $|\tilde{V}_G|$. Cycle 6 shows pronounced hysteresis, indicating that the detrapping threshold lies between −47.7 and −52.7 V (≈ −48.2 V; fig. S3); no significant hysteresis is observed in cycles 1-5 confirming that the threshold $V_G$ is not exceeded at lower $|\tilde{V}_G|$.

Within the zero-resistance (superconducting) regime (purple-shaded region in Fig. 2A, "$R = 0$"), $I_{leak}$ and $I_c$ are anticorrelated: a local decrease in $|I_{leak}(t)|$ within a certain $V_G$ range is accompanied by a corresponding increase in $I_c(t)$ (black arrows, Fig. 2A). Once $|I_{leak}|$ overcomes a certain threshold, $I_c$ drops to zero, marking the transition to the resistive (normal) state (gray-shaded region, "$R \neq 0$"). The $|I_{leak}|(|V_G|)$ curves in Fig. 2B further show that positive-polarity sweeps are systematically shifted toward higher $|V_G|$ compared to negative-polarity sweeps, which provides direct evidence of net electron trapping.

Analysis of multiple devices shows that the onset of non-monotonic behavior in $I_{leak}(V_G)$ or $I_c(V_G)$ marks the threshold $V_G$ above which a significant number of charges become trapped or detrapped. To determine this threshold precisely, we initialized Device F in the electron-trapped state and applied successive $V_G$ cycles, with progressively increasing $|\tilde{V}_G|$. Below the threshold $V_G$ (~ 48 V for this device) hysteresis is negligible; once exceeded, pronounced hysteresis develops (cycle 6, Fig. 2, C and D). The threshold $V_G$ therefore defines the minimum $V_G$ required to modify the charge-trapping state and thus needed to perform WRITE and ERASE operations ($V_{WRITE}$ and $V_{ERASE}$, respectively).

To further verify the charge-trapping dynamics, pulsed-sequence measurements were also performed on Device F (Supplementary Text and fig. S3). Consistent with the dependence of trapping on $V_G$ polarity, Device B shows no evidence of charge trapping when $V_G$ sweeps are restricted to a single $V_G$ polarity at fixed $\tilde{V}_G$ (fig. S5).

Across all devices tested, we find that the threshold $V_G$ varies by less than a factor of two despite substantial variations in other parameters such as $I_{c0}$ or gate-to-wire distance (Table S2). For all investigated devices, the threshold $V_G$ for significant charge trapping is much higher than the minimum $V_G$ at which $I_c$ suppression (i.e., the GCS) occurs. This separation is what enables non-destructive readout of the charge-trapping state via $I_c$ measurements at an appropriate $V_G$ for reading, $V_{READ}$, which can be chosen $\approx -22$ V for Device F (Fig. 2D).

**Non-volatile charge-trap memory based on GCS**

The hysteretic charge-trapping behavior established above provides the basis for a voltage-controlled, non-volatile superconducting memory. We demonstrate memory operation using Devices D and I (Fig. 3), although we reproduced memory functionality across all devices tested to this purpose (Devices F to I, Supplementary Text).

Memory operation relies on two coupled dependencies: (i) $I_c$ is determined by the power dissipated by the gate $P_G = V_G \cdot I_{leak}$ (Fig. 3A, bottom panel), and (ii) $P_G$ is controlled by the charge-trapping state. The first dependency has been established in previous GCS studies – $I_c$ suppression in GCS devices is uniquely determined by $P_G$, independent of the device history (*18,20,23,24*). The second dependency is the new element introduced here: charge trapping modifies $I_{leak}$ at fixed $V_G$, thereby shifting $P_G$ and, consequently, $I_c$.

Figure 3A illustrates the operating principle using Device D. After a WRITE operation, electron trapping shifts $I_{leak}(V_G)$ toward higher (positive) $V_G$ (middle panel). At fixed $V_G$, this reduces $I_{leak}$ and hence $P_G$, which in turn increases $I_c$ – which manifests as a shift of $I_c(V_G)$ toward higher $V_G$ (top panel). We define the electron-trapped state as state "1" (red, after $V_{WRITE}$) and the charge-detrapped state as state "0" (blue, after $V_{ERASE}$). Since the $I_c(P_G)$ dependence is unchanged by charge trapping (bottom panel), the two states correspond to two distinct operating points on the same $I_c(P_G)$ curve – state "1" at lower $P_G$ and higher $I_c$, and state "0" at higher $P_G$ and lower $I_c$ – yielding two well-separated reproducible $I_c$ levels that encode the binary memory states.

Reversible memory operation is demonstrated on Device I through repeated sequences of READ, WRITE and ERASE operations (Fig. 3B). A positive pulse $V_{WRITE}$ = + 76.2 V switches the device from state "0" to state "1", and a negative pulse $V_{ERASE}$ = -73.8 V restores state "0". Readout is performed at $V_{READ}$ = 56.0 V, a voltage large enough to produce well-separated $I_c$ levels between the two states yet below the trapping threshold – which ensures non-destructive readout.

Figure 3C shows 49 consecutive WRITE-READ-ERASE cycles on Device I, each following the sequence in Fig. 3B. The $I_c$ values extracted from $V(I)$ characteristics remain positive throughout all operations, confirming that the device always stays in its zero-resistance (superconducting) state except during transient switching events. The unperturbed $I_{c0}$ measured at $V_S = 0$ before the cycling sequence is also shown as reference (black dot, Fig. 3C).

Figure 3D shows a zoom of the READ operations from Fig. 3C and reveals two distinct $I_c$ populations: state "1" ($I_1$, red) and state "0" ($I_0$, blue), both below $I_{c0}$ (black line), as expected from the shift in the operating point. The corresponding $I_{leak}$ levels, $I_{leak,1}$ (red) and $I_{leak,0}$ (blue), are shown in Fig. 3E, and the $P_G$ dissipated during READ is shown in Fig. 3F. On average, $P_G$ is of ~0.3 nW for state "1" (red) and of ~1.2 nW for state "0" (blue). For comparison, WRITE and ERASE operations dissipate ~20 nW and ~50 nW, respectively, in the same device – higher, as expected, since charge trapping or detrapping must be activated. We emphasize that the READ operation is non-destructive and that the device remains in the zero-resistance state during readout except for transient switching events.

The memory is truly non-volatile: the stored state is preserved throughout thermal cycling well above $T_c$ and back (fig. S7), confirming that the information is encoded in trapped charges rather than in the superconducting condensate. Moreover, WRITE operations can be performed in the normal state (e.g., at $T$ = 50 K; fig. S8), demonstrating that information storage does not require the device to be in the superconducting state. This combination of thermal robustness and above-$T_c$ writability is absent from all existing superconducting memory technologies and represents a key distinguishing feature of the present approach.

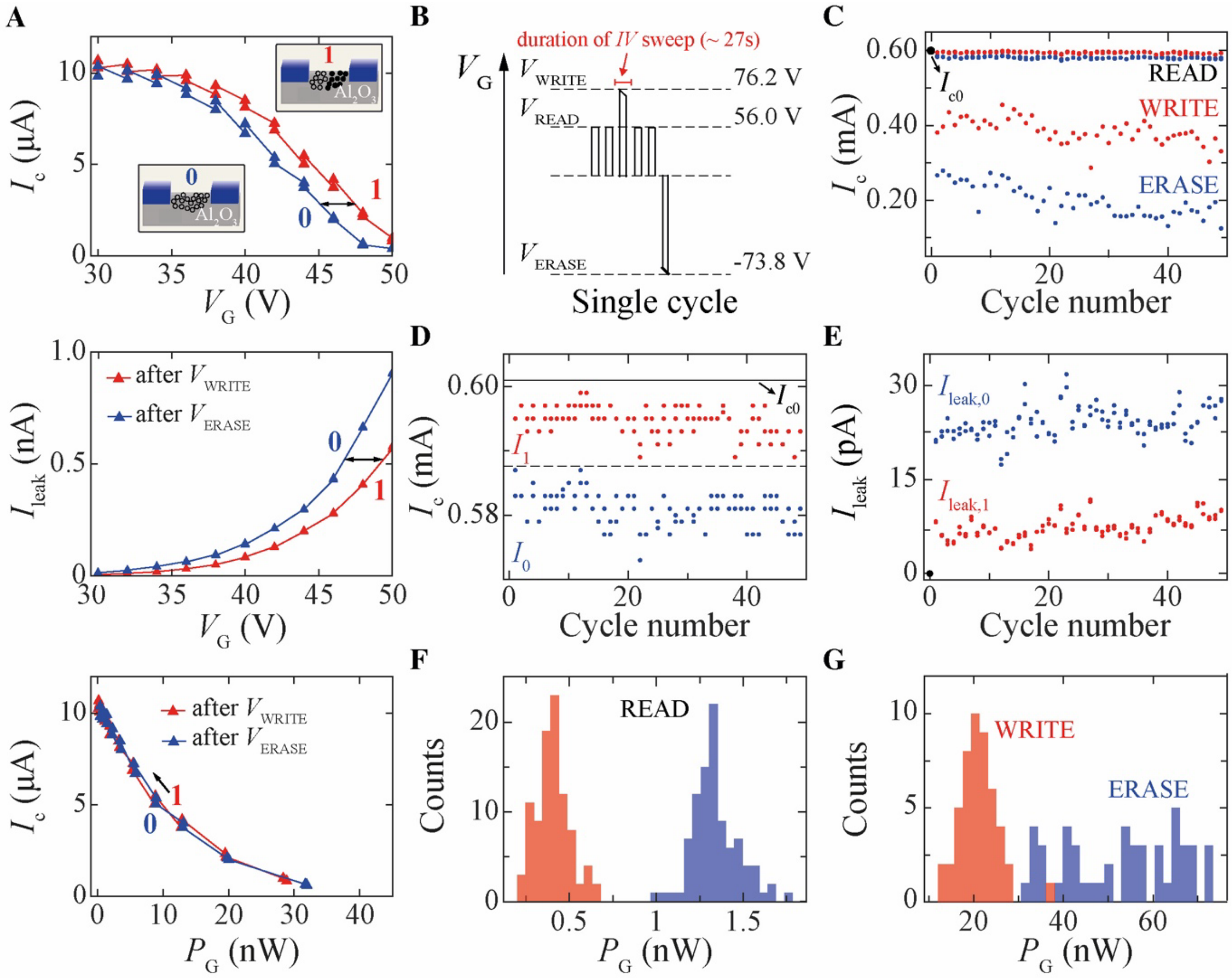

**Fig. 3. Non-volatile charge trap memory device based on GCS.** (**A**) Critical current $I_c$ versus gate voltage $V_G$, $I_c(V_G)$, (top), leakage current versus $V_G$, $I_{leak}(V_G)$, (middle), and $I_c$ versus gate power, $I_c(P_G)$, (bottom) measured on Device D after WRITE (state "1", red) and ERASE (state "0", blue) operations. Both states follow the same $I_c(P_G)$ scaling, with state "1" shifted toward lower $P_G$ (arrow) due to reduced $I_{leak}$ after charge trapping. (**B** to **G**) Memory operation demonstrated on Device I via voltage versus current, $V(I)$, sweeps at fixed sourced $V_G$. (B) Schematic of one complete WRITE-READ-ERASE cycle. Before the first cycle, a $V(I)$ sweep at $V_S = 0$ establishes the reference critical current $I_{c0}$. A total of 49 cycles are performed, each labeled by a cycle number. (C) $I_c$ versus cycle number for all operations: WRITE (red dots), ERASE (blue dots), and READ (red or blue dots following each WRITE or ERASE, respectively). The reference $I_{c0}$ is shown as a black dot. Two well-separated $I_c$ levels are maintained across all 49 cycles. (D) Zoom of the READ $I_c$ values from (C) showing two stable populations $I_1$ (state "1", red) and $I_0$ (state "0", blue) as a function of cycle number. (E) Corresponding $I_{leak}$ levels $I_{leak,1}$ (red) and $I_{leak,0}$ (blue) versus cycle number. (F) Histogram of gate power $P_G = V_{READ} \cdot I_{leak}$ dissipated during READ operations with average values of ~ 0.3 nW (state "1") and ~ 1.2 nW (state "0"). (G) Same as (F) for WRITE (~ 20 nW) and ERASE (~ 50 nW) operations.

## Integration of GCS memory cells into a NAND architecture and pathways to optimization

Having established the operation of individual memory cells, we now consider their integration into the standard NAND architecture used in CMOS charge-trap memories. Figure 4A illustrates the conventional hierarchy of a NAND layout: a die contains planes, each subdivided into blocks of pages, where a page consists of individually addressable memory cells. A block is the smallest erasable unit, while word lines select pages and bit lines address individual cells.

In what follows, we focus on the READ operation in a GCS-based NAND architecture and its advantages over CMOS charge-trap technology. WRITE and ERASE operations are discussed in the Supplementary Text.

Figure 4B shows a representative memory block with eight voltage-addressable word lines (pages, horizontal) and *N* bit lines (vertical, labelled 1 to *N*). Each bit line is connected to a bit-line selection transistor and a ground (GND) selection transistor, which can be implemented using CMOS field-effect transistors or superconducting equivalents such as GCS elements or nTrons (*29,36*). Throughout this discussion, we assume all transistors and interconnects to be non-resistive (superconducting), and each memory cell to consist of a GCS charge-trap device with two states, ”1” and ”0” (blue rectangles in Fig. 4B).

As in CMOS NAND, a $V_{\mathrm{READ}}$ is applied to the selected word line during readout (Fig. 4B, orange rectangle). However, a key distinction from CMOS is the elimination of the pass voltage ($V_{\mathrm{pass}}$): in conventional CMOS NAND, $V_{\mathrm{pass}}$ must be applied to all unselected (idle) cells to maintain them in a conductive state (*37*), which gives a substantial contribution to standby power dissipation. In the GCS NAND architecture, unselected cells are instead held at ground potential and remain in their non-resistive (zero-resistance) state, eliminating $V_{\mathrm{pass}}$-induced dissipation entirely. Residual standby power dissipation in idle cells arises only from $I_{\mathrm{leak}}$ through the gate dielectric.

The READ operation is illustrated in Fig. 4B. To read a target cell, its bit line is enabled via the corresponding selection transistor (light blue boxes, Fig. 4B), while the GND selection transistors remain on. A constant bias current $I$ that satisfies $I_0 < I < I_1 \approx I_{c0}$ is injected through the selected bit line toward GND (fig. S10 and Supplementary text), where $I_1$ and $I_0$ are the $I_c$ of the states “1” and “0”, respectively. The small series resistor $R_{\mathrm{bias}}$ (Fig. 4B) is the only dissipative element in the bit line and should be minimized to reduce dissipated power.

When the selected cell is in state “0”, its critical current $I_0$ is lower than the applied $I$, which causes the memory device to switch to its resistive state. This transition can occur, for example, via hotspot formation induced by $I_{\mathrm{leak}}$, similarly to the switching mechanism in superconducting nanowire single-photon detectors (SNSPDs) under photon illumination (*38-40*). Provided that $R_{\mathrm{N}} > R_{\mathrm{bias}}$, the bias current $I$ is then redirected through $R_{\mathrm{bias}}$ to GND, giving a measurable voltage signal that identifies state “0” (option 2 in Fig. 4, B and C). To minimize power dissipation and prevent latching of the full bit line, bit lines should have low kinetic inductance (*40*) and the selection transistor should be turned off immediately after readout.

When the selected memory cell is in state “1”, $I$ remains below $I_1$ and the memory device remains in its non-resistive state, so that the bias current $I$ flows without dissipation through bit line to GND (option 1 in Fig. 4, B and C). The absence of a voltage drop across $R_{bias}$ identifies state "1". This dissipationless readout of state "1" is a direct consequence of the zero-resistance nature of the superconducting memory cell and has no equivalent in CMOS technology. A practical advantage of this READ scheme is that all cells on a selected page can be read out simultaneously, since $V_{READ}$ is applied to the entire word line. The memory state can be determined using either a current sensor or a voltage sensor: a current sensor distinguishes both state "1" and state "0" (options 1 and 2 in Fig. 4, B and C), whereas a voltage sensor detects only state "0" via the voltage drop across $R_{bias}$.

Figure 4D shows a cross-sectional schematic of an optimized GCS charge-trap memory cell incorporating a dedicated charge-trapping layer (orange) between gate and dielectric layers (gray) – an alternative to the single-dielectric geometry used in this work that is expected to improve trapping efficiency and reduce operating voltages. An overview of further GCS cell geometries is provided in Supplementary Text.

The READ scheme of Figure 4 can be extended beyond charge-trap devices to GCS cells employing high-impedance memristive dielectrics (e.g., $TiO_x$-based devices (*41*)). Although the storage mechanism in memristive dielectrics differs from charge trapping, the general operation principle of the memory remains governed by the dependence of $I_c$ on $P_G$.

Although the $V_{WRITE}$ and $V_{ERASE}$ voltages of our current GCS memory devices are relatively large, they are comparable in magnitude to those of fin field-effect transistors (FinFETs) (*42*) and can be substantially reduced by engineering the gate dielectric. Specifically, replacing the $Al_2O_3$ dielectric – which hosts deep charge traps that require a high $V_G$ to fill (*43*) – with a thin top-gate dielectric containing shallower traps closer to the interface with the S should bring operational voltages in the range (< 5 V) compatible with existing cryogenic CMOS devices.

Using a pulse-sequence method, we have also demonstrated WRITE and ERASE operations with durations of 500 ms over more than 50 cycles (Devices F and H; figs. S3 and S4). These timescales, however, are limited by our measurement setup bandwidth rather than by the intrinsic device dynamics. READ operation (tested at $V_{READ}$ ~ 20 V for Device F) are similarly bandwidth-limited in our setup to the millisecond range.

Much faster operation is achievable in principle. WRITE/ERASE times in the μs range should be achievable by using a top-gated geometry and a measurement setup with larger bandwidth, as already shown for SiN-based charge-trap memories (*44*). GCS devices embedded in resonators can respond to $V_G$ changes on timescales of few ns (*45*) – this is the equivalent to a single READ event. In the Supplementary Text, we also discuss how $V_{READ}$ can be defined after a brief initial characterization of $I_{c0}$.

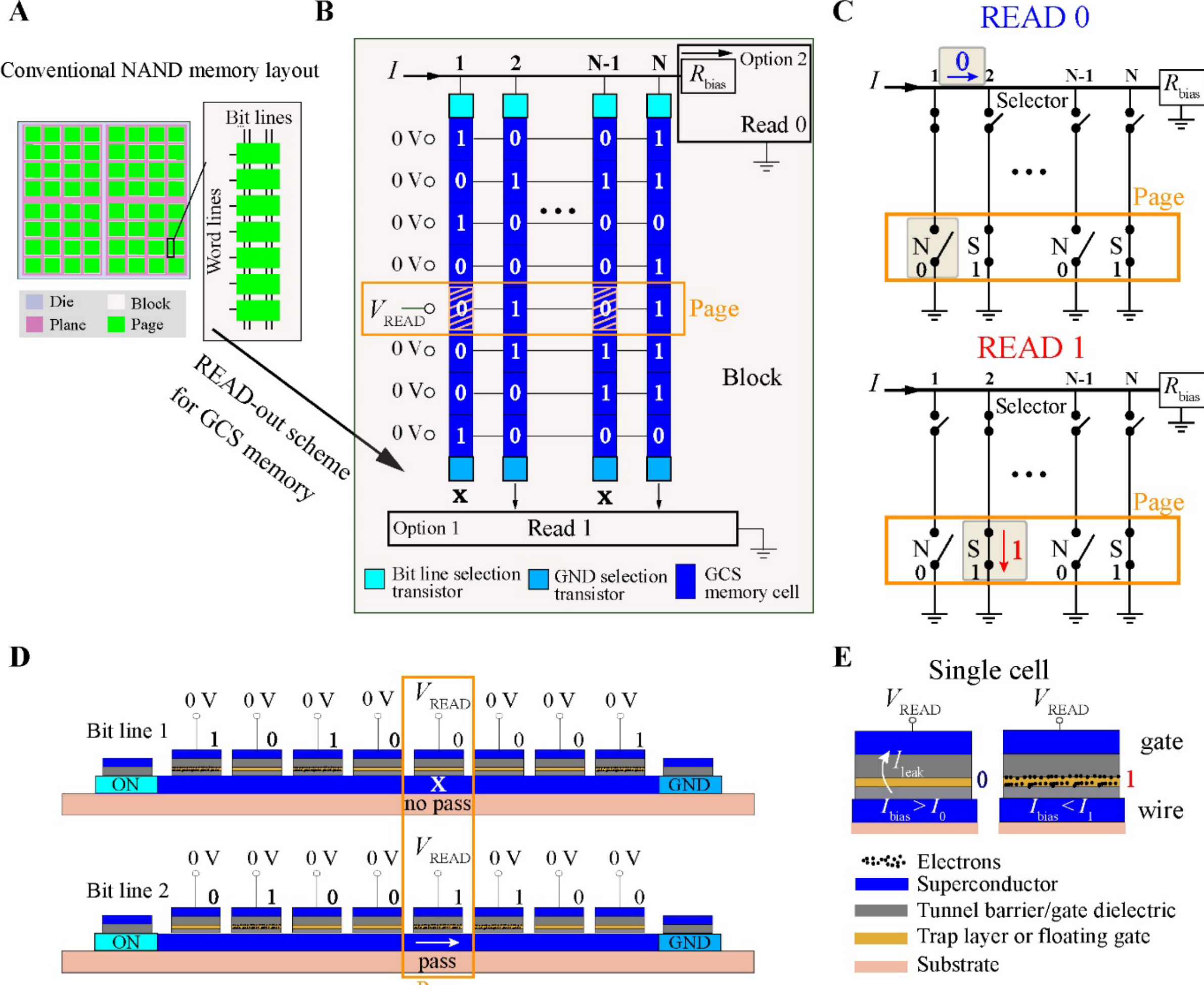

**Fig. 4. Integration of GCS memory cells into a NAND architecture.** (**A**) Schematic of the conventional NAND memory hierarchy: a die contains planes, subdivided into blocks of pages, with each page comprising individually addressable memory cells. The block is the smallest erasable unit. (**B**) READ operation in a GCS charge-trap NAND array. A read voltage $V_{\mathrm{READ}}$ is applied to the selected word line (orange rectangle) and a bias current $I_0 < I < I_1$ is injected along each bit line ($I_0$ and $I_1$ being the critical current $I_c$ of state "0" and "1", respectively). Depending on the cell's memory state, $I$ either flows without dissipation through the bit line to GND (state "1", option 1) or is redirected through the low-ohmic resistance $R_{\mathrm{bias}} < R_{\mathrm{N}}$ (state "0", option 2), with $R_{\mathrm{N}}$ being the normal-state resistance of the memory device. All GND selection transistors are turned on during the READ operation. (**C**) Simplified scheme of the READ operation for the selected page in (B) showing options 1 and 2 for states "1" and "0", respectively. (**D**) Cross-sectional schematic of the memory cells on bit lines 1 and 2 from (B) showing a top-gated geometry with a dedicated charge-trapping layer (orange) between gate dielectrics (gray). (**E**) Operating principle of the two memory states. In state "0", no charges are trapped, $I_{\mathrm{leak}}$ between the gate and the wire is large, which increases $P_{\mathrm{G}}$ and in turn reduces $I_c$ to $I_0 < I$, switching the cell to the resistive state. In state "1", charges are trapped within the charge-trapping layer, which reduces the effective $V_{\mathrm{G}}$ between gate and the wire, suppressing $I_{\mathrm{leak}}$ and $P_{\mathrm{G}}$, so that $I_c = I_1 > I$ and the cell remains in the zero-resistance state.

For a NAND array with current-pulse-based readout, sub-ns READ times appear feasible: analogous readout schemes have already achieved sub-ns timing in SNSPD arrays (*46,47*), and the underlying physics – a current pulse-triggered normal-state transition – is identical to the readout of the state "0" described above.

As in mature CMOS charge-trap memories, adopting a top-gated architecture with controlled dielectric growth is expected to substantially improve device-to-device reproducibility of the threshold $V_G$ – a key requirement for large-scale NAND integration. Strategies to reduce power consumption and optimize the operating point of single devices are discussed in the Supplementary Text, together with an outlook on further device improvements and remaining challenges to address.

Power dissipation in a GCS NAND array depends on the choice of superconducting material, gate dielectric, device geometry, sensing scheme, and the value of $R_{bias}$. By combining charge-trapping dielectrics with energetically shallow traps with superconductors that exhibit $I_c$ suppression at low $P_G$, readout dissipation could in principle be reduced to the pW range for state "0" and the fW range for state "1" — orders of magnitude below CMOS equivalents. To achieve READ speeds comparable to or exceeding those of CMOS flash memories, superconductors with low kinetic inductance and short coherence length — such as NbN or NbTiN — are the most promising candidates.

## CONCLUSIONS

In this work, we establish a new approach to superconducting memory by combining the GCS with charge-trapping in a dielectric substrate, which fills a longstanding technological gap in superconducting electronics, where memories comparable to CMOS memories have long been sought.

The key mechanism underlying our memory functionality is the coupling between charge trapping in a dielectric substrate and the gate power $P_G = V_G \cdot I_{leak}$ governing the GCS. Charge trapping shifts the $I_{leak}(V_G)$ characteristics, thereby modifying $P_G$ at fixed $V_G$ and enabling two stable states with distinct $I_c$ values suitable for reliable information storage.

Through nearly fifty consecutive WRITE, ERASE and READ operations, we demonstrate robust, non-destructive readout and reversible switching of the memory states, with the device remaining in the zero-resistance state throughout all operations except transient switching events during state "0" readout. Stored information survives thermal cycling well above $T_c$ and can be written in the normal (non-superconducting) state — capabilities that are absent from all existing superconducting memory technologies. The combination of non-volatility, voltage control, thermal robustness, and dissipationless readout represents a paradigm shift in the design space of superconducting memories.

Integration of GCS charge-trap memory cells into a NAND architecture offers significant advantages over CMOS charge-trap flash: unselected cells require no $V_{pass}$ and remain in the zero-resistance state, reducing standby dissipation to the sub-nW level, while state "1" readout is entirely dissipationless. With optimized dielectric materials and top-gated geometries, operating voltages are projected to fall into the few-volt range and READ times into the sub-ns regime — competitive with, or exceeding, state-of-the-art CMOS flash.

These results establish GCS charge-trap memory as a compelling platform for cryogenic computing and offer a route toward the seamless integration of superconducting logic and memory in future high-performance and quantum computing systems.

## Materials and Methods

**Sample fabrication.** All Nb devices with a Ti adhesion layer were deposited onto commercial single-crystalline (1-102)-oriented $Al_2O_3$ substrates (CrysTec manufacturer). These one-side polished substrates were diced into pieces of 5x5 mm with thickness of 530 μm.

Before device patterning and material deposition, the $Al_2O_3$ substrates were cleaned through a process involving first ultrasonication in acetone and then in isopropanol (for ~ 5 minutes each), followed by a blow-dry with pure $N_2$. Subsequently, a layer of polymethyl methacrylate (950 PMMA A4 from Kayaku) was spun onto the cleaned substrates and baked for 90 seconds at 180 °C on a hot plate. After cooling the substrates to room temperature, a conductive polymer, Electra 92 (AR-PC 5092.02, Allresist), was spun onto the PMMA as top layer.

The device patterning was then carried out using a single e-beam lithography (EBL) step with a 20 kV acceleration voltage. The exposure dose used was ~ 300 μC/cm². After beam exposure, the conductive top coating was removed by rinsing the substrate in deionised (DI) water. To this purpose, the sample was dipped in a first beaker and then in a second beaker, both containing fresh DI water, for ~ 1 min, followed by a dry blow of $N_2$. The mask (positive tone) was then developed by dipping the sample into a methylisobutylketone (MIBK) solution consisting of 3 parts of IPA with 1 part of MIBK for 25 s. The fully-developed samples were then loaded into an ultra-high vacuum (UHV) chamber (base pressure $< 2 \times 10^{-8}$ Torr) and pumped for at least 12 hours to allow the polymer masks to degas.

In the UHV chamber, a Ti adhesion layer was sputtered by radiofrequency (RF) magnetron sputtering (200 W, 17sccm Ar flow, deposition pressure 1.5 mTorr). Right after Ti adhesion layer growth, a Nb (S) layer was deposited at 300 W using simultaneously 2 RF magnetron guns and one direct current (DC) magnetron gun, with an Ar flow rate of 17 sccm and a deposition pressure of 1.5 mTorr. Before deposition, the sputter rates for Ti and Nb in the same growth conditions had been calibrated using atomic force microscopy, yield values of 0.04 nm/s for Ti and 0.3 nm/s for Nb.

After deposition, the Nb devices were placed in a 50 °C hot acetone bath for at least 3h to perform lift-off, followed by ultrasonication for ~90 seconds to remove the unwanted material. The devices were then cleaned in IPA and dried using $N_2$. Finally, before wedge bonding the samples to a printed circuit board (PCB) for low-temperature transport measurements, the samples were stored in either $N_2$ or under vacuum to prevent oxidation of the Nb layer.

**Transport measurements.** Resistance versus temperature, $R(T)$, measurements and voltage versus current, $V(I)$, characteristics were measured inside a dry inverted cryostat (Dry ICE 3K

INV) using a standard 4-point measurement configuration. For current bias, a low-noise Keithley 6221 current source was used, and a nanovoltmeter (Keithley 2182A) was employed to measure the voltage drop across the Dayem bridge. Throughout the measurements, the compliance limit for $I_{leak}$ was set to 10.5 nA.

The measurement setup is equipped with two-stage RC filters on all lines to minimise noise interference. These filters consist of a series resistance of ~2 kΩ and a capacitance of 4 nF. The total resistance of the wires, $R_{wiring}$, is 2.05(1) kΩ. The temperature during measurements was carefully controlled to maintain a bath temperature within ± 10 mK or better. A low-noise source-measure unit (Keithley 6430) in conjunction with a pre-amplifier was used in a two-wire configuration to source the gate voltage $V_G$ and measure simultaneously $I_{leak}$. To minimise parasitic leakage currents, four additional shielded and unfiltered lines (resistance between 1 and 2 Ω) were installed in the systems. These lines feature a TΩ resistance to ground under an applied voltage of 70 V and at room temperature and guarantee further isolation of the measurement setup from external noise .

$V_G$-dependent measurements were performed using a stepwise ramp for the sourced voltage $V_S$. At each step of the ramp, $V_S$ was held constant while either a series of $V(I)$ sweeps were done to determine the corresponding critical current $I_c$ and $I_{leak}$ (see, e.g., middle and bottom panels of Fig. 3A of the main text). The effective $V_G$ was determined using the formula $V_G = V_S - R_{wiring} \cdot I$. After measuring a series of $V(I)$ characteristics and the corresponding $I_c$ and $I_{leak}$, the applied $V_S$ was ramped to the next value. This protocol ensures that each extracted $I_c$ corresponds to a well-defined and stationary condition of the applied gate voltage.

## Acknowledgements

We thank Jennifer Koch for helpful discussions and Alfredo Spuri for setting up the measurement program. We acknowledge the use of the experimental equipment and the expert support concerning its usage provided by the nano.lab at the University of Konstanz. All authors also acknowledge funding from the EU's Horizon 2020 research and innovation program under Grant Agreement No. 964398 (SUPERGATE) and from the University of Konstanz via a Zukunftskolleg research fellowship.

## Contributions

L. R. designed the experiment, which was supervised by A.D.B. and E.S. L.R. executed the measurements and the data analysis and discussed the experimental data with A.D.B. and E.S. All authors contributed to the writing of the manuscript and commented on it.

## Competing interests

The authors declare no competing interests.

## Data availability

The datasets generated and analyzed during the current study are available from the corresponding authors on reasonable request.

## References

1. T. Hiramoto, Five nanometre CMOS technology. *Nat. Electron.* **2**, 557–558 (2019). https://doi.org/10.1038/s41928-019-0343-x
2. V. Natarajan, A. Deshpande, S. Solanki, A. Chandrasekhar, Thermal and power challenges in high performance computing systems. *Jpn. J. Appl. Phys.* ***48***, 05EA01 (2009). https://doi.org/10.1143/JJAP.48.05EA01
3. D. S. Kalashnikov, V. I. Ruzhitskiy, A. G. Shishkin, I. A. Golovchanskiy, M. Y. Kupriyanov, I. I. Soloviev, D. Roditchev, V. S. Stolyarov, Demonstration of a Josephson vortex-based memory cell with microwave energy-efficient readout. *Commun. Phys.* **7**, 88 (2024). https://doi.org/10.1038/s42005-024-01570-4
4. B. Baek, W. H. Rippard, S. P. Benz, S. E. Russek, P. D. Dresselhaus, Hybrid superconducting-magnetic memory device using competing order parameters. *Nat. Commun.* **5**, 3888 (2014). https://doi.org/10.1038/ncomms4888
5. Y. Takeshita, F. Li, D. Hasegawa, K. Sano, M. Tanaka, T. Yamashita, A. Fujimaki, High-speed memory driven by SFQ pulses based on 0- π SQUID. *IEEE Trans. Appl. Supercond.* **31**, 1100906 (2021). https://doi.org/10.1109/TASC.2021.3060351
6. S. Alam, M. S. Hossain, S. R. Srinivasa, A. Aziz, Cryogenic memory technologies. *Nat. Electron.* **6**, 185–198 (2023). https://doi.org/10.1038/s41928-023-00930-2
7. Y. Polyakov, S. Narayana, V. M. Semenov, Flux trapping in superconducting circuits. *IEEE Trans. Appl. Supercond.* **17**, 520-525 (2007). https://doi.org/10.1109/TASC.2007.898707
8. S. Alam, M. S. Hossain, K. Ni, V. Narayanan, A. Aziz, Voltage-controlled cryogenic Boolean logic gates based on ferroelectric SQUID and heater cryotron. *J. Appl. Phys.* **135**, 014903 (2024). https://doi.org/10.1063/5.0172531
9. E. M. Spanton, M. Deng, S. Vaitiekėnas, P. Krogstrup, J. Nygård, C. M. Marcus, K. A. Moler, Current-phase relations of few-mode InAs nanowire Josephson junctions. *Nat. Phys.* **13**, 1177-1181 (2017). https://doi.org/10.1038/nphys4224
10. F. Vigneau, R. Mizokuchi, D. C. Zanuz, X. Huang, S. Tan, R. Maurand, S. Frolov, A. Sammak, G. Scappucci, F. Lefloch, S. De Franceschi, Germanium quantum-well Josephson field-effect transistors and interferometers. *Nano Lett*. **19**, 1023-1027 (2019). https://doi.org/10.1021/acs.nanolett.8b04275
11. M. Nadeem, M. S. Fuhrer, X. Wang. The superconducting diode effect. *Nat. Rev. Phys*. **5**, 558-577 (2023). https://doi.org/10.1038/s42254-023-00632-w
12. L. Bauriedl, C. Bäuml, L. Fuchs, C. Baumgartner, N. Paulik, J. M. Bauer, K.-Q. Lin, J. M. Lupton, T. Taniguchi, K. Watanabe, C. Strunk, N. Paradiso, Supercurrent diode effect and magnetochiral anisotropy in few-layer $NbSe_2$. *Nat. Commun*. **13**, 4266 (2022). https://doi.org/10.1038/s41467-022-31954-5
13. H. Wu, Y. Wang, Y. Xu, P. K. Sivakumar, C. Pasco, U. Filippozzi, S. S. P. Parkin, Y.-J. Zeng, T. McQueen, M. N. Ali, The field-free Josephson diode in a van der Waals heterostructure. *Nature* **604**, 653-656 (2022). https://doi.org/10.1038/s41586-022-04504-8
14. M. Gupta, G. V. Graziano, M. Pendharkar, J. T. Dong, C. P. Dempsey, C. Palmstrøm, V. S. Pribiag, Gate-tunable superconducting diode effect in a three-terminal Josephson device. *Nat. Commun*. **14**, 3078 (2023). https://doi.org/10.1038/s41467-023-38856-0
15. A. Paghi, L. Borgongino, E. Strambini, G. De Simoni, L. Sorba, F. Giazotto, The ferroelectric superconducting field effect transistor (2025), pre-print available at https://arxiv.org/abs/2507.04773.
16. O. Medeiros, M. Castellani, V. Karam, R. Foster, A. Simon, F. Incalza, B. Butters, M. Colangelo, K. K. Berggren, A scalable superconducting nanowire memory array with row-column addressing. *Nat. Electron*. **9**, 69-77 (2026). https://doi.org/10.1038/s41928-025-01512-0.
17. G. De Simoni, F. Paolucci, P. Solinas, E. Strambini, F. Giazotto, Metallic supercurrent field-effect transistor, *Nat. Nanotechnol*. **13**, 802-805 (2018). https://doi.org/10.1038/s41565-018-0190-3
18. L. Ruf, E. Scheer, A. Di Bernardo, High-performance gate-controlled superconducting switches: large output voltage and reproducibility. *ACS Nano* **18**, 20600-20610 (2024). https://doi.org/10.1021/acsnano.4c05910
19. M. F. Ritter, A. Fuhrer, D. Z. Haxell, S. Hart, P. Gumann, H. Riel, F. Nichele, A superconducting switch actuated by injection of high-energy electrons. *Nat. Commun*. **12**, 1266 (2021). https://doi.org/10.1038/s41467-021-21231-2

20. T. Elalaily, O. Kürtössy, Z. Scherübl, M. Berke, G. Fülöp, I. E. Lukács, T. Kanne, J. Nygård, K. Watanabe, T. Taniguchi, P. Mall, S. Csonka, Gate-controlled supercurrent in epitaxial Al/InAs nanowires. *Nano Lett*. **21**, 9684-9690 (2021). https://doi.org/10.1021/acs.nanolett.1c03493
21. L. D. Alegria, C. G. L. Bøttcher, A. K. Saydjari, A. T. Pierci, S. H. Lee, S. P. Harvey, U. Vool, A. Yacoby, High-energy quasiparticles injection into mesoscopic superconductors. *Nat. Nanotech*. **16**, 404-408 (2021). https://doi.org/10.1038/s41565-020-00834-8
22. I. Golokolenov, A. Guthrie, S. Kafanov, Y. A. Pashkin, V. Tsepelin, On the origin of the controversial electrostatic field effect in superconductors. *Nat. Commun*. **12**, 2747 (2021). https://doi.org/10.1038/s41467-021-22998-0
23. J. Basset, O. Stanisavljević, M. Kuzmanović, J. Gabelli, C. H. L. Quay, J. Estève, M. Aprili, Gate-assisted phase fluctuations in all-metallic Josephson junctions. *Phys. Rev. Res*. **3**, 043169 (2021). https://doi.org/10.1103/PhysRevResearch.3.043169
24. T. Jalabert, E. F. C. Driessen, F. Gustavo, J. L. Thomassin, F. Levy-Bertrand, C. Chapelier, Thermalization and dynamic of high-energy quasiparticles in a superconducting nanowire. *Nat. Phys.* **19**, 956-960 (2023). https://doi.org/10.1038/s41567-023-01999-4
25. J. Koch, C. Cirillo, S. Battisti, L. Ruf, Z. M. Kakhaki, A. Paghi, A. Gulian, S. Teknowijoyo, G. De Simoni, F. Giazotto, C. Attanasio, E. Scheer, A. Di Bernardo, Gate-controlled supercurrent effect in dry-etched Dayem bridges of non-centrosymmetric niobium rhenium. *Nano Res.* **17**, 6575-6581 (2024). https://doi.org/10.1007/s12274-024-6576-7
26. M. F. Ritter, N. Crescini, D. Z. Haxell, M. Hinderling, H. Riel, C. Bruder, A. Fuhrer, F. Nichele, Out-of-equilibrium phonons in gated superconducting switches. *Nat. Electron.* **5**, 71–77 (2022). https://doi.org/10.1038/s41928-022-00721-1
27. H. Du, Z. Xu, Z. Wei, D. Li, S. Chen, W. Tian, P. Zhang, Y.-Y. Lyu, H. Sun, Y.-L. Wang, High-energy electron local injection in top-gated metallic superconductor switch. *Supercond. Sci. Technol.* **36**, 095005 (2023). https://doi.org/10.1088/1361-6668/ace65f
28. C. Puglia, G. De Simoni, F. Giazotto, Electrostatic control of phase slips in Ti Josephson nanotransistors. *Phys. Rev. Applied* **13**, 054026 (2020). https://doi.org/10.1103/PhysRevApplied.13.054026
29. L. Ruf, C. Puglia, T. Elalaily, G. De Simoni, F. Joint, M. Berke, J. Koch, A. Iorio, S. Khorshidian, P. Makk, S. Gasparinetti, S. Csonka, W. Belzig, M. Cuoco, F. Giazotto, E. Scheer, A. Di Bernardo, Gate control of superconducting current: Mechanisms, parameters, and technological potential. *Appl. Phys. Rev.* **11**, 041314 (2024). https://doi.org/10.1063/5.0222371
30. T. Elalaily, M. Berke, I. Lilja, A. Savin, G. Fülöp, L. Kupás, T. Kanne, J. Nygård, P. Makk, P. Hakonen, S. Csonka, Switching dynamics in Al/InAs nanowire-based gate-controlled superconducting switch. *Nat. Commun*. **15**, 9157 (2024). https://doi.org/10.1038/s41467-024-53224-2
31. T. Mikolajick, M. Sprecht, N. Nagel, T. Mueller, S. Riedel, F. Beug, T. Melde, K.-H. Kusters, The Future of Charge Trapping Memories. *Proceedings VLSI-TSA*, 130 (2007). https://doi.org/10.1109/VTSA.2007.378943
32. L. Manchanda, W. H. Lee, J. E. Bower, F. H. Baumann, W. L. Brown, C. J. Case, R. C. Keller, Y. O. Kim, E. J. Laskowski, M. D. Morris, R. L. Opila, P. J. Silverman, T. W. Sorsch, G. R. Weber, Gate quality doped high films for CMOS beyond 100 nm: 3-10 nm $Al_2O_3$ with low leakage and low interface states. *Tech. Dig. Int. Electron. Devices Meet.* 605-608 (1998). https://doi.org/10.1109/IEDM.1998.746431
33. S. Sambuco Salomone, J. Lipovetzky, S. H. Carbonetto, M. A. García Inza, E. G. Redin, F. Campabadal, A. Faigón, Experimental evidence and modeling of two types of electron traps in $Al_2O_3$ for non-volatile memory applications. *J. Appl. Phys*. **113**, 074501 (2013). https://doi.org/10.1063/1.4792038
34. M. Specht, H. Reisinger, M. Stadele, F. Hofmann, A. Gschwandtner, E. Landgraf, R. J. Luyken, T. Schulz, J. Hartwich, L. Dreeskornfeld, W. Rosner, J. Kretz, L. Risch, Retention time of novel charge trapping memories using $Al_2O_3$ dielectrics. *33rd Conference European Solid-State Device Research*, 155-158 (2003). https://doi.org/10.1109/ESSDERC.2003.1256834
35. E. Schilirò, R. Lo Nigro, P. Fiorenza, F. Roccaforte, Negative charge trapping effects in $Al_2O_3$ films grown by atomic layer deposition onto thermally oxidized 4H-SiC. *AIP Adv*. **6**, 075021 (2016). https://doi.org/10.1063/1.4960213

36. A. N. McCaughan, K. K. Berggren, A superconducting-nanowire three-terminal electrothermal device, *Nano Lett.* **14**, 5748-5753 (2014). https://doi.org/10.1021/nl502629x
37. Y.-H. Hsiao, H.-T. Lue, W.-C. Chen, K.-P. Chang, B.-Y. Tsui, K.-Y. Hsieh, C.-Y. Lu, Impact of $V_{pass}$ Interference on Charge-Trapping NAND Flash Memory Devices. *IEEE Trans. Device Mater. Reliab.* **15**, 136-141 (2015). https://doi.org/10.1109/TDMR.2015.2398193.
38. C. M. Natarajan, M. G. Tanner, R. H. Hadfield, Superconducting nanowire single-photon detectors: physics and applications. *Supercond. Sci Technol*. **25**, 063001 (2012). https://doi.org/10.1088/0953-2048/25/6/063001
39. A. D. Semenov, G. N. Gol'tsman, A. A. Korneev, Quantum detection by current carrying superconducting film. *Phys. C: Supercond. Appl*. **351**, 349-356 (2001). https://doi.org/10.1016/S0921-4534(00)01637-3
40. A. J. Annunziata, O. Quaranta, D. F. Santavicca, A. Casaburi, L. Frunzio, M. Ejrnaes, M. J. Rooks, R. Cristiano, S. Pagano, A. Frydman, D. E. Prober, Reset dynamics and latching in niobium superconducting nanowire single-photon detectors. *J. Appl. Phys.* **108**, 084507 (2010). https://doi.org/10.1063/1.3498809
41. Y. Beilliard, F. Paquette, F. Brousseau, S. Ecoffey, F. Alibart, D. Drouin, Investigation of resistive switching and transport mechanisms of $Al_2O_3/TiO_{2-x}$ memristors under cryogenic conditions (1.5 K). *AIP Adv*. **10**, 025305(2020). https://doi.org/10.1063/1.5140994
42. J.-H. Ahn, D.-II Moon, S.-W. Ko, C.-H. Kim, J.-Y. Kim, M.-S. Kim, M.-L. Seol, J.-B. Moon, J.-M. Choi, J.-S. Oh, S.-J. Choi, Y.-K. Choi, A SONON device with a separate charge trapping layer for improvement of charge injection. *AIP Adv*. **7**, 035205 (2017). https://doi.org/10.1063/1.4978322
43. D. Liu, S. J. Clark, J. Robertson, Oxygen vacancy levels and electron transport in $Al_2O_3$. *Appl. Phys. Lett.* **96**, 032905 (2010). https://doi.org/10.1063/1.3293440
44. J. Hur, D. Kang, D.-I. Moon, J.-M. Yu, Y. K. Choi, S. Yu, Cryogenic storage memory with high-speed, low-power, and long-retention performance. *Adv. Electron. Mater*. **9**, 2201299 (2023). https://doi.org/10.1002/aelm.202201299
45. F. Joint, K. R. Amin, I. P. C. Cools, S. Gasparinetti, Dynamics of gate-controlled superconducting bridges. *Appl. Phys. Lett*. **125**, 092602 (2024). https://doi.org/10.1063/5.0222058
46. M. Ejrnaes, A. Casaburi, R. Cristiano, N. Martucciello, F. Mattioli, A. Gaggero, R. Leoni, J.-C. Villégier, S. Pagano, Characterisation of superconducting pulse discriminators based on parallel NbN nanostriplines. *Supercond. Sci. Technol.* **24**, 035018 (2011). https://doi.org/10.1088/0953-2048/24/3/035018
47. A. J. Kerman, D. Rosenberg, R. J. Molnar, E. A. Dauler, Readout of superconducting nanowire single-photon detectors at high count rates. *J. Appl. Phys*. **113**, 144511 (2013). https://doi.org/10.1063/1.4799397
48. K. Matsunaga, T. Tanaka, T. Yamamoto, Y. Ikuhara Y, First-principles calculations of intrinsic defects in $Al_2O_3$. *Phys.Rev. B* **68**, 085110 (2003). https://doi.org/10.1103/PhysRevB.68.085110
49. Y. Miura, Y: Matukura, Investigation of silicon-silicon dioxide interface using MOS structure. *Jpn. J. Appl. Phys.* **5**, 180 (1966). https://doi.org/10.1143/JJAP.5.180
50. K. O. Jeppson, C. M. Svensson, Negative bias stress of MOS devices at high electric fields and degradation of MNOS devices. *J. Appl. Phys*. **48**, 2004-2014 (1977). https://doi.org/10.1063/1.323909
51. A. Hiraiwa K. Horikawa, H. Kawarada, M. Kado, K. Danno, Influence of $Al_2O_3$ atomic-layer deposition temperature on positive-bias instability of metal/$Al_2O_3$/$\beta$-$Ga_2O_3$ capacitors. *J. Vac. Sci. Technol. B* **42**, 012207(2024). https://doi.org/10.1116/6.0003186
52. T. Grasser, Stochastic charge trapping in oxides: From telegraph noise to bias temperature instabilities. *Microelectron. Reliab.* **52**, 39-70 (2012). https://doi.org/10.1016/j.microrel.2011.09.002
53. T. Grasser, H. Reisinger, W. Goes, Th. Aichinger, Ph. Hehenberger, P.-J. Wagner, M. Nelhiebel, J. Franco, B. Kaczer, Switching oxide traps as the missing link between negative bias temperature instability and random telegraph noise. *Proc. Int. Electron. Devices Meeting* (IEDM), 729-732 (2009). https://doi.org/10.1109/IEDM.2009.5424235
54. Z. Li, M. Sotto, F. Liu, M. K. Husain, H. Yoshimoto, Y. Sasago, D. Hisamoto, I. Tomita, Y. Tsuchiya, S. Saito, Random telegraph noise from resonant tunneling at low temperatures. *Sci Rep*. **8**, 250 (2018). https://doi.org/10.1038/s41598-017-18579-1
55. J. Yao, L. Zhong, D. Natelson, J. M. Tour, In situ imaging of the conducting filament in a silicon oxide resistive switch. *Sci. Rep.* **2**, 242 (2012). https://doi.org/10.1038/srep00242

56. Z. Hou, Q. Zhang, H. Yin, J. Xiang, C. Qin, J. Yao, J. Gu, Fabrication and characterisation of p-channel charge trapping type FOI-FinFET memory with MAHAS structure. *ECS J. Solid State Sci. Technol.* **6**, Q136 (2017). https://doi.org/10.1149/2.0251710jss
57. A. D. Inglis, Y. Le Page, P. Strobel, C. M. Hurd, Electrical conductance of crystalline $Ti_nO_{2n-1}$ for n = 4-9. *J. Phys. C: Solid State Phys.* **16**, 317 (1983). https://doi.org/10.1088/0022-3719/16/2/015.